\documentclass[submission,copyright,creativecommons]{eptcs}
\providecommand{\event}{ICLP/LPNMR-DC 2024} 

\usepackage{iftex}
\usepackage{amssymb} 

\ifpdf
  \usepackage{underscore}         
  \usepackage[T1]{fontenc}        
\else
  \usepackage{breakurl}           
\fi

\title{Logical Foundations of Smart Contracts}
\author{Kalonji Kalala
\institute{School of Electrical Engineering and Computer Science\\
University of Ottawa\\
Ottawa, Canada}
\email{hkalo081@uottawa.ca}
}
\def\titlerunning{Logical Foundations of Smart Contracts}
\def\authorrunning{Kalonji Kalala }
\begin{document}
\maketitle

\begin{abstract}
Nowadays, sophisticated domains are emerging which require appropriate formalisms to be specified accurately in order to reason about them. One such domain is constituted of smart contracts that have emerged in cyber physical systems as a way of enforcing formal agreements between components of these systems. 
Smart contracts self-execute to run and share business processes through Blockchain, in decentralized systems, with many different participants. Legal contracts are in many cases complex documents, with a number of exceptions, and many subcontracts. The implementation of smart contracts based on legal contracts is a long and laborious task, that needs to include all actions, procedures, and the effects of actions related to the execution of the contract.
An ongoing open problem in this area is to formally account for smart contracts using a uniform and somewhat universal formalism. This thesis proposes logical foundations to smart contracts using the Situation Calculus, a logic for reasoning about actions.
Situation Calculus is one of the prominent logic-based artificial intelligence approaches that provides enough logical mechanism to specify and implement dynamic and complex
systems such as contracts. Situation Calculus is suitable to show how worlds dynamically change.  Smart contracts are going to be implement with Golog (written en Prolog), a Situation Calculus-based programming language for modeling complex and dynamic behaviors.
\end{abstract}

\section{Introduction}
This work is motivated by the increasing amount of investigations around Blockchains~\cite{MonratSA19} and cyber physical systems~\cite{PanarelloTM0P18} that are coupled with the necessity of smart contracts. Cyber physical systems are smart devices that are connected together with the purpose of collecting data, processing them, and providing intelligent decisions. A Blockchain is represented as a public ledger  that requires a software to be shared by peers and run a business. That  shared software representing a legal contract among parties is called smart contract. The implementation of a legal smart contract needs to take in account all the complexity of a contract and its numerous subcontracts. There are many existing approaches used to implement smart contracts. Those approaches present a number of limitations in handling all the complexity of smart contracts and their large number of subcontracts. To solve those limitations, the need of an approach to represent legal smart contracts in a way that allows to verify their correctness becomes evident. Logic-based approaches are emerging to provide mechanisms for reasoning about actions and their effects. 

This thesis is using the Situation Calculus, a logic for specifying dynamical systems in artificial intelligence, to specify smart contracts and reason about them. Many logic-based formalisms have been investigated previously to represent smart contracts,  such as the Deontic Logic \cite{Azzopardi2018}, Event Calculus \cite{KruijffW19}, Defeasible Logic \cite{GiannikisD06} and other logic-based approaches in \cite{abs-2008-02712}. 
The idea of specifying smart contracts in the Situation Calculus is mentioned for the first time in \cite{DBLP:journals/ais/DaskalopuluS97}. Daskalopoulou's main idea is that monitoring and enforcing smart contracts can be supported in a particular application domain by giving a suitable (and in formal view) representation for agreements between parties that accounts for deviations of the parties’ behavior from their obligations and corrections of such deviations. Hence a form of Deontic Logic combined with a dynamic, temporal logic is advocated for modeling smart contracts. She uses Event Calculus to this end, while mentioning that the Situation Calculus could as well have been used for this purpose.  Our thesis embarks on a program of incorporating obligation-producing actions into Situation Calculus to capture and specify smart contracts. 

\subsection*{Smart Contracts}

A {\it smart contract (e-contract)}  is a piece of software that monitors the correct execution of the legal contracts \cite{DBLP:journals/corr/abs-2104-03764}\cite{surden2012computable}\cite{DBLP:journals/access/RahmanRHHAG19}\cite{Daskalopulu2019}.  A smart contract is a  automation of the execution of an agreement.
An application's rules and regulations can be digitally facilitated, verified, validated, and enforced using a smart contract, which is an  executable code on a blockchain network. Without the involvement of other parties, smart contracts enable legitimate transactions. These transactions can be monitored and maintained irreversible \cite{ramamurthy2020blockchain}. For blockchain applications\cite{DBLP:journals/access/RahmanRHHAG19}\cite{elsden2018making}, a smart contract fulfils the need for application-specific verification and validation. A blockchain is a new form of infrastructure that has the potential to profoundly alter the way that individuals transact, communicate, organize, and identify themselves \cite{zheng2018blockchain}\cite{han2023accounting}\cite{xu2023survey}\cite{mourtzis2023blockchain}. A blockchain system is composed of a network of computational nodes that share a single data structure (the blockchain) and reach agreement on its current state \cite{dorri2017blockchain}\cite{yue2016healthcare}.
\subsection*{Smart Contract Formalization}
Formal languages are mostly involved in the specification of e-contracts. Both syntax and semantics of formal languages are beneficial in the process of either verificating or validating of e-contrats \cite{DBLP:journals/internet/KrishnaK08}\cite{DBLP:journals/ail/GovernatoriIMRS18}.  Furthermore, le formal languages are equipped to handle business vocabularies and rule semantics. To enable users to comprehend high-level logic and accurately express e-contract semantics, effective approaches to solve logic-specification consistency must be developed. Many investigations have been carried out to study the use of different formal languages in the modeling of electronic contracts, formal languages such as Event, Default, Situation and Deontic calculus. 

\section{Problem Statement}\label{STATEMENT}

An ongoing open problem in this area of smart contracts is to formally account for them using a uniform and somewhat universal formalism. In other words, the challenge is to give a formal semantics for smart contracts. This thesis proposes logical foundations to smart contracts using the Situation Calculus. Since smart contracts deal with obligations of contracting agents, the problem statement amounts to formalizing the notion of obligation in the Situation Calculus, and subsequently using this formalization to specify smart contracts, execute those specifications and use these specifications to prove properties of the smart contracts. 

\section{Situation Calculus}\label{SITUATIONCALCULUS}

The situation calculus \cite{mccarthy1963situations,DBLP:conf/birthday/Reiter91} is a many-sorted and mostly first order language with equality specifically designed for representing dynamically changing world. We consider a version of the situation calculus with four sorts for actions, situations, time points, and objects other that the first three sorts. {\bf Actions} are first order terms consisting of an action function symbol and its arguments, one of which being the action occurrence time. {\bf Situations} are first order terms denoting finite sequences of actions. They are represented using a binary function symbol $do$: $do(\alpha,s)$ denotes the sequence resulting from adding the action      $\alpha$ to an existing sequence $s$. The constant $S_0$ ({\em initial situation}) denotes the empty sequence $[\;]$.   {\bf Time points} are the sequence of real numbers. Finally, {\bf objects} represent domain specific individuals other than actions, situations, and time points.  The language has an alphabet with variables and a finite number of constants for each sort, a finite number of function symbols called {\it action functions}, a finite number of function symbols called {\it situation independent functions}, a finite number of function symbols called {\it functional fluents}, a finite number of predicate symbols called {\it situation independent predicates}, and a finite number of predicate symbols called {\it predicate fluents}. Predicate fluents represent properties whose truth values vary from situation to situation as a consequence of executions of  actions. A predicate fluent is denoted by a predicate symbol whose last argument is a situation term. Functional fluents denote values that  vary from situation to situation as a consequence of executions of actions.  
The language  also  includes special predicates $Poss$, and $\sqsubset$; $Poss(a,s)$ means that the action $a$ is possible in the situation $s$, and $s \sqsubset s'$ states that the situation $s'$ is reachable from $s$ by performing some sequence of actions. 
In contract modelling terms, $s \sqsubset s'$ means that $s$ is a proper 
subtrace of the contract execution $s'$.  
The predicate $\sqsubset$ will be useful in formulating properties of 
contracts.  
\\ \\
The Situation Calculus allows a high level of flexibility in representing dynamic environments, thus making possible flexible and nuanced modeling of real-world scenarios. The formalism of Situation Calculus offers an effective way to describe changes and reason about the consequences of actions inside a system. This makes it possible to express change and action in an efficient manner. In our case, a smart contract represents the complex dynamic domain representing the Situation Calculus.
\\

A dynamic  domain (e.g.,  legal contracts) is axiomatized in the Situation Calculus with axioms which describe how and under what conditions the domain is changing or not changing as a result of performing actions. Such axioms are called {\it basic action theory} in~\cite{reiter2001knowledge}. They include the following classes of sentences: domain independent foundational axioms for situations; action precondition axioms, one for each action term, stating the conditions of change; successor state axioms, one for each fluent, stating how change occurs; specific axioms for time, stating the occurrence times of actions and start times of situations;  unique names axioms for action terms; and axioms describing the initial situation of the domain.   
In addition to the primitive actions mentioned above, complex actions mimicking Algol-like programming language constructs have been introduced to capture the full expressiveness of application domains. These complex actions are going to be implement with GOLOG \cite{mcilraith2001adapting}\cite{hofmann2016continual}, a Situation Calculus-based programming language for complex and dynamic behaviors. Golog interpreter is written in Prolog.

\section{Solution}\label{SOLUTION}

Since smart contracts deal with obligations of contracting agents, our approach consists in formalizing the notion of obligation in the Situation Calculus. In doing so, we extend a well-known solution by Scherl and Levesque \cite{scherl2003knowledge} to the classical frame problem for knowledge to obligations. We then use the Situation Calculus with obligations to specify smart contracts as follows.  First, we construct logical theories called {\it basic contractual theories} to formalize legal contracts. Basic contractual theories provide the formal semantics of the corresponding legal contracts. Second, we represent legal contracts as processes in the Situation Calculus; such processes lead to situations where desirable properties hold that logically follow from the basic contractual theory representing those legal contracts. We provide an implementable specification, thus allowing one to automatically check many properties of the specification using an interpreter.  We use the interpreter to develop a framework for obligation-based programming which we apply to verify properties of the specified smart contracts.

\section{Methodology}

In this thesis, we based our research on a methodology proposed in Reiter's version of the Situation Calculus~\cite{reiter2001knowledge, DBLP:conf/birthday/Reiter91} by representing smart contract transaction as Situation Calculus actions whose effects are captured as truth values of Situation Calculus fluents. This approach relies on modelling a smart contract as a mostly first-order theory called {\it basic action theory} ~\cite{reiter2001knowledge}, augmented with sentences that account for the embedding of the logic of obligations, the so-called deontic logic, into the Situation Calculus. The resulting theory is the {\it basic contractual theory} mentioned in Section~\ref{SOLUTION}. 
In addition to four foundational axioms for situations \cite{reiter1993proving,pirri1999some} that structure the space of situations, basic contractual theories contain a set of {\it successor state axioms} which extend Reiter's solution to the frame problem to the embedding of deontic logic into the situation calculus.  With the basic contractual theories in hand, we represent smart contracts in Situation Calculus as  logic-based programs built  using complex actions that macro-expand to a sequence of simple actions. These programs are executed using an interpreter that uses the basic contractual theories as background theories for the purpose of proving properties of the formalized smart contracts.   

\section{Expected Contributions and Goals}

The overall goal of the research that underpins the thesis has been stated in Section~\ref{STATEMENT}, namely providing logical foundations that constitute a formal semantics for smart contracts; and, to this end, we choose the Situation Calculus as our logic. The expected specific contributions and goals include the following: 
\begin{itemize}
  \item Exploring existing logic-based approaches for formalizing smart contracts and comparing them. 
   \item Extending the solution by Scherl and Levesque to the classical frame problem for knowledge to obligations in the Situation Calculus.  
   \item Representing complex actions for specifying smart contracts as such actions. 
  \item Specifying and proving properties of smart contracts in the language of the Situation Calculus augmented with obligations.  
  \item Extending the GOLOG interpreter~\cite{DBLP:journals/jlp/LevesqueRLLS97},     a situation calculus-based programming language for defining complex actions in terms of a set of primitive actions axiomatized in the situation calculus. 
  \item Using GOLOG to develop a framework for obligation-based programming.  
  \item Providing a Prolog implementation of the framework.  
\end{itemize}



\section{Current Status of the Research}

With reference to sections below steps (2)--(4) and parts of Step (5) have been completed: 
\begin{enumerate}
    \item {\bf Background and Related Work}. Here, we introduce the various definitions of smart contract concepts, the ontology and related  formalizations of legal contracts. We introduce the different logical formalisms used in modeling smart contracts, such as Event Calculus, Default Logic, Modal Logic, Deontic Logic and Temporal Logic. Also, we present some important related works on logic-based smart contracts. Finally, we compare the presented  logical formalisms based on a number of features.
    \item {\bf Formal Preliminaries}. In this chapter, we present the situation calculation as enriched by Reiter in~\cite{reiter2001knowledge}. We present the language of the Situation Calculus,  its syntax, and  its foundational axioms. We then present the main components of the situation calculus machinery, including the basic actions theories and the regression mechanism used for  reasoning about actions.  Furthermore, we present Scherl and Levesque's extension of the Situation Calculus to account for knowledge and knowledge-producing actions that explicitly create agent knowledge. Scherl and Levesque's solution to the frame problem for knowledge-producing actions is presented, as this solution forms the departure point of our own solution for accounting for obligations. Finally we summarize the main theoretical and practical results of the Situation Calculus.
    \item {\bf Formalization of Obligations}. This chapter  extends the  solution by Scherl and Levesque to the frame problem for knowledge-producing actions to obligation-producing actions. An obligation-producing action is one which enacts obligations on the part of whoever agent performs it.  Both works have their roots in the seminal work of Raymond Reiter who proposed the so-called successor state axioms as a solution to the frame problem for actions that are neither knowledge, nor obligation producing.  The specification in this chapter yields intuitive properties that one would expect from obligations.  Obligation-producing actions do only affect a newly introduced fluent for capturing the notion of obligation in the Situation Calculus, and no other fluents, except those fluents that are made obligatory by obligation-producing actions. In addition, persistence appears as a consequence of this new setting: if something is obligatory to an agent in a given situation, it remains obligatory  everywhere as it should be, unless something contrary to the obligation occurs. We show that Reiter's regression operator for reasoning about actions back to the initial situation is a reasoning mechanism for this setting as well.
    
    \item {\bf Formalization of Smart Contracts}. We use the Symboleo \cite{sharifi2020symboleo} ontology to specify smart contracts.  Symboleo is defined as a formal specification language for legal contracts\cite{DBLP:conf/er/ParvizimosaedSA20}. Based on Event Calculus, it contains axioms to specify its semantics and its syntax is formalized by a grammar \cite{Sharifi2020}. Symboleo generates smart contracts from natural language contract. In order to  make an actual legal contract larger, Symboleo includes a contract ontology containing components such as : obligations, powers and state models to apply to the concepts of contracts \cite{Parvizimosaed2020}. We will use the Symboleo ontology and language to formulate smart contract Symboleo programs which will be systematically translated to Situation Calculus specifications in an effort to have implementable specifications.
    
    \item {\bf Obligation-Based Programming Framework}. This could be addressed by extending of GOLOG to an Obligation-Based Programming Framework and writing the Prolog Implementation of the Programming Framework.

\end{enumerate}

	
 
\bibliographystyle{eptcsini}
\bibliography{generic}

\end{document}